\documentclass[twocolumn]{revtex4}

\usepackage{amsmath}    
\usepackage{graphicx}   
\usepackage{verbatim}   
\usepackage{color}      
\usepackage{subfigure}  
 
\def\lowsim{\mathrel{\lower 0.7 ex \hbox to 0 pt{$\sim$\hss}}}

 \newcommand{\wpi}{\omega_{pi}}
 \newcommand{\wci}{\omega_{ci}}

 \newcommand{\m}[1]{\mathbf #1} 

\begin{document}

\title{Dependence of Kinetic Plasma Turbulence on Plasma beta}
\author{Tulasi N. Parashar}
\author{William H. Matthaeus}
\author{Michael A. Shay}
\affiliation{Bartol Research Institute, Department of Physics and Astronomy, University of Delaware, Newark, DE}

\begin{abstract}
We study the effects of plasma $\beta$ (ratio of plasma pressure to magnetic pressure) on the evolution of kinetic plasma turbulence using fully kinetic particle-in-cell simulations of decaying turbulence. We find that the plasma $\beta$ systematically 
affects spectra, measures of intermittency, decay rates of  turbulence fluctuations, and  partitioning over different channels of energy exchange More specifically, an increase in plasma $\beta$ leads to greater total heating, with proton heating preferentially more than electrons. Implications for achieving magnetosheath like temperature ratios are discussed.
\end{abstract}

\maketitle
\section{Introduction}
Turbulence at kinetic scales in weakly collisional plasmas such as the solar wind is a complex interplay of small amplitude turbulent fluctuations and their interaction with particle distributions. In most cicumstances, energy at the large magnetohydrodynamic (MHD) scales is cascaded via nonlinear interactions to kinetic scales where potentially many plasma effects become important, in what may be called the range of kinetic cascade(s) \citep[e.g.][]{SchekochihinApJS09,YangPRE17}. At these kinetic scales, the interactions between turbulent fluctuations and particles transfer energy through a multiplicity of channels \cite{YangEA17ab}, leading eventually to a net transfer into internal energy (or {\it random}) degrees of freedom. 

The solar wind (as well as the Earth's magnetosheath) shows variability in various characteristic plasma parameters including notably the plasma $\beta$ (the ratio of thermal to magnetic pressure). This is broadly expected based on elementary plasma theory given that the propeties of waves, damping rates and instabilities in linear Vlasov theory generally show significant dependence on $\beta$. A question naturally arises then as to how the nature of solar wind turbulence varies with $\beta$, other parameters being fixed.
Although we use the solar wind as our principle example, one also finds a range of values of $\beta$ in other venues in the heliosphere. For example the solar corona below the Alfv\'en critical point (or, region) almost certainly resides at $\beta < 1$ including most likely some very low values. Moving outward, the first $\beta = 1$ surface is reached around 30 $R_\odot$, although this occurs potentially at much lower altitudes near the heliospheric current sheet \cite{ChhiberEA18}. In the terrestrial magnetosheath the ram pressure of the shocked solar wind can raise the $\beta$ to values substantially greater than unity. Throughout most of the magnetosphere, however, the plasma $\beta$ tends to be less than unity. On the other hand, in the outer heliosphere, where pickup ions of interstellar origin form a halo at ten or more thermal speeds away from the core distribution\cite{IsenbergJGR87}, the effective $\beta$ may be very high. Similar large ranges of plasma $\beta$ may be found in other astrophysical contexts such as radiatively inefficient accretion flows (RIAFs) \cite{QuataertApJ98, QuataertAN03, SharmaApJ03}. Consequently, the effect of varying $\beta$ on plasma turbulence is a matter of broad importance. 
We address that question here using a direct strategy based on numerical plasma simulations.

In the past decade or so, with the ongoing advances in computational capabilities, there have been numerous advances in understanding plasma turbulence. Studies have focused on examining turbulence properties for various parameters, varying turbulence amplitude, the direction and strength of the mean magnetic field (guide field), the dimensionality, and the plasma beta, as well as numerical parameters such as proton-electron mass, hybrid model electric field closure, speed of light, grid resolution and system size. Each of these has important effects on simulation results, and it would be counterproductive to progress to claim that any of these factors may be neglected. This complexity is a major reason that the problem of kinetic turbulence is very challenging.

The effects of variations of plasma beta have been studied in the literature for many years, with older works (too numerous to review in detail) typically concentrating on one dimensional problems involving waves, instabilities, or manifestations of linear damping rates in turbulent regimes (see e.g. \cite{QuataertApJ98, GaryBook, HowesMNRAS10}). With recent advances \cite{BowersPP08} multidimensional problems and simulation of far-from-equilibrium or turbulent plasmas have become more commonplace \cite{KarimabadiPP13, RoytershteynPTRSA15}, with much of the state of the art work devoted to applications such as fusion plasmas, and specialized models such as gyrokinetics or fluid-electron hybrid PIC models \cite{KarimabadiPP14}. In this vast literature there are numerous examples of studies that have varied plasma beta to examine specific effects such as variation in spectra in hybrid simulations \cite{VasquezApJ12}, effect of change in electron thermal physics \cite{ParasharPP14} and spectral breaks vs $\beta$ from simulations and space data [e.g. \cite{ChenGRL14, FranciApJ16, WangApJ18, WangJGR18,woodham2018role}. Another issue is the distribution of plasma states in the  parameter space spanned by parallel plasma beta and temperature anisotropy, an issue that has been examined extensively in observations \cite{BalePRL09, HellingerGRL06, MatteiniGRL07, MarucaPRL11, MarucaApJ12} and in simulations \cite{ServidioApJL14, ServidioJPP15}. We have not found existing results that provide a systematic quantitative account of the influences of variation of beta upon the characteristics of fully kinetic plasma turbulence such as decay and heating rates, apportionment of these into proton and electron internal energies, intermittency, etc. 

Here we study beta variation effects by carrying out a set of full-electromagnetic kinetic PIC simulations of moderate system size, with all parameters, including initial turbulence properties, fixed with the sole exception of the plasma beta. No approximations other than artificial mass ratio, speed of light, and the 2.5D geometry have been made. The restriction to 2.5D geometry affords the possibility of high spatial resolution and large system size, enabling  examination of high effective Reynold number turbulence \cite{ParasharApJ15}. Meanwhile the restriction to 2.5D does not greatly affect the qualitative nature of heating, its distribution with current density \cite{KarimabadiPP13, KarimabadiSSR13, LiApJL16, WanPP16}, or the distribution of coherent structures \cite{GrecoApJL09}.The focused purpose of the paper is to document how variation in $\beta$ affects the properties of turbulence and the kinetic features such as heating and anisotropy. 

\section{Kinetic Simulations}
To address this question, we perform fully kinetic PIC simulations of a fully ionized proton-electron plasma using the code P3D \cite{ZeilerJGR02}. The code has been extensively used for space plasma applications of reconnection and turbulence. The simulations are performed in a 2.5D setup with turbulent fluctuations in $(x,y)$ plane but no variation in the $z$ direction. All physical quantities have all 3 components of the vectors. All the results are presented in normalized units where lengths are normalized to $d_i=c/\wpi$, the ion inertial length, where $c$ is the speed of light, and $\wpi$ is the plasma frequency, time is normalized to $\wci^{-1}$, inverse of the proton cyclotron frequency, and speeds to the Alfv\'en speed based on normalization magnetic field $V_{A0}=B_0/\sqrt{4\pi n_0}$ where $B_0$ is the arbitrary normalizing magnetic field and $n_0$ is the normalizing density. The simulation domain is $L=149.6 d_i$. The grid has $N_x=N_y=4096$ points, 3200 particles per cell of each species with 107 billion total particles. $dx=\lambda_e$ (for the $\beta=0.6$ case), $m_e/m_i=0.04$, density $n=1 n_0$, $\beta_i =\beta_e =0.3,0.6,1.2$. $\beta$ values much smaller or much larger than these make fully kinetic PIC simulations computationally much more expensive and are out of the scope of this study. Varying ratio of proton/electron $\beta$ is also left for future studies. 

The initial conditions for all simulations are exactly the same with rms $\delta b = 0.5B_0$, $\delta v = 0.5 V_{A_0}$, except that plasma $\beta$ is varied by adjusting the initial (equal) temperatures of protons and electrons. This means that the turbulent Mach number varies with simulations; particularly the sonic mach number decreases with increasing $\beta$. The initial conditions are created by populating Fourier modes with $2 \le |k| \le 4$ with random phased fluctuations and a specific spectrum. This is typical of the ``Alfv\'enic'' initial conditions used in decaying simulations of MHD turbulence. Here too, the system evolves without external forcing. Each run is evolved for more than $300 \wci^{-1}$. Next we discuss the properties of turbulence as the system evolves.

\section{Results } 

\begin{figure}[!hbt]
\begin{center}
\includegraphics[width=3.0in]{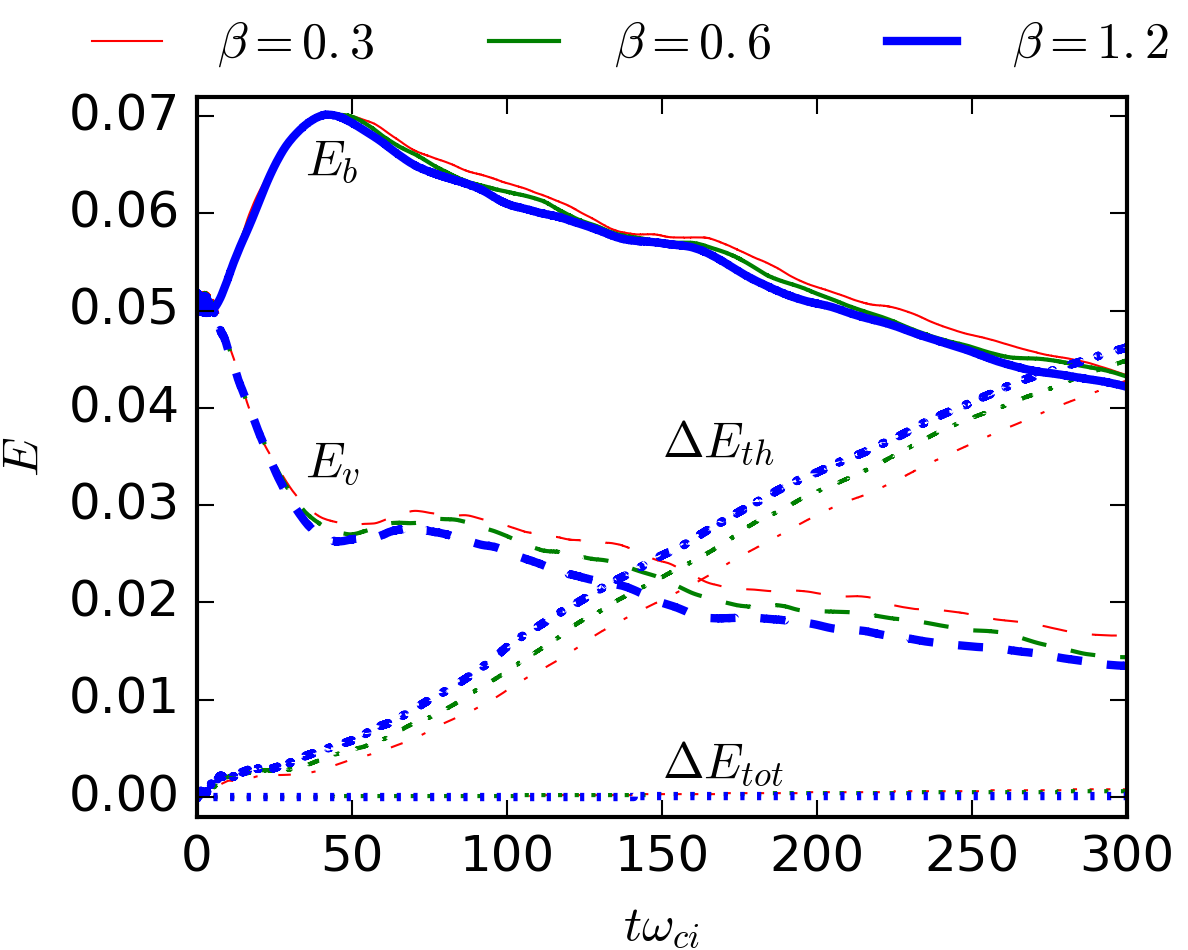}
\caption{(Color Online) Fluctuations energies, magnetic $E_b$ and flow $E_v$, 
as well as the change in internal energy $\Delta E_{th}$ for 
simulations with plasma $\beta = $0.3, 0.6 and 1.2, 
thicker lines denoting larger $\beta$.  Change in total energy is also shown
to verify excellent energy conservation.
A dependence on $\beta$ is seen in the time evolution. 
Turbulent fluctuations decay more for larger $\beta$, 
as more energy goes into the internal energy.
 }
\label{energetics}
\end{center}
\end{figure}

\begin{figure}[!hbt]
\begin{center}
\includegraphics[width=3.0in]{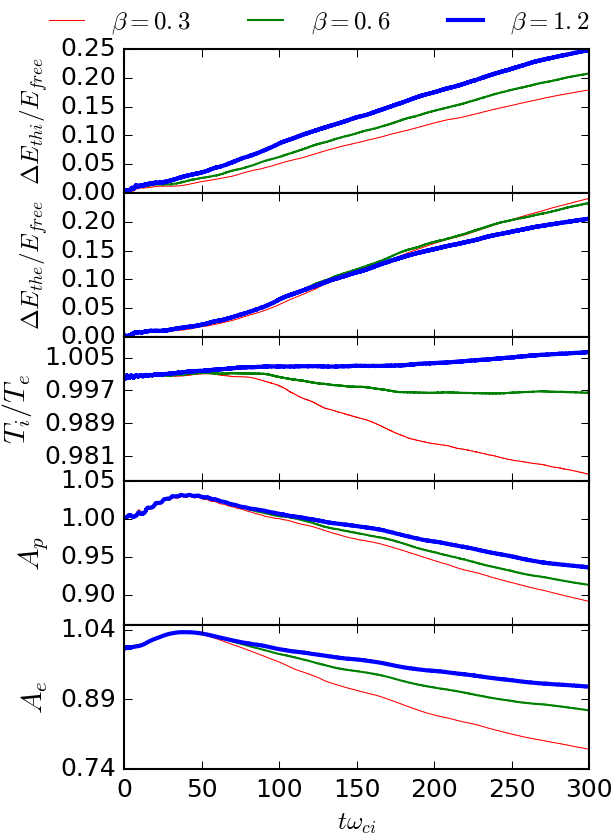}
\caption{(Color Online) Proton and electron heating as a function of plasma $\beta$. Colors represent $\beta$, solid lines represent protons and dashed lines represent electrons. Proton internal energy increases with $\beta$ whereas the change in electron internal energy gets smaller. Middle panel shows the ratio of proton and electron heating, and bottom two panels show the anisotropy $A=T_\perp/T_\parallel$ for each species.
 }
\label{heating}
\end{center}
\end{figure}

As a first step, we examine the energetics of the simulations. Figure \ref{energetics} shows the time evolution of magnetic, flow, and thermal energies for each simulation. The flow and thermal energies are summed over species. It is evident that the loss of turbulent energy and gain of internal energy increases with increasing $\beta$. Therefore we see immediately that the overall heating rate increases with $\beta$. To arrive at this conclusion, some care has been taken to design the simulations so that the total energy is reasonably well-conserved. Typically the runs conserved total energy (not including the energy content of the uniform applied mean magnetic field) to better than 1\% and $\Delta E_{tot} <<$ any fluctuation energy of interest. See Figure \ref{energetics}.

At early times, $\omega_{ci} t < 40$, the flow and magnetic fields go through a typical Alfv\'enic exchange, and this phase of the evolution shows minimal, if any, dependence upon $\beta$. This time coincides with a few estimated nonlinear times. Evidently the early relaxation from the random phased initial conditions to a less-stressed state, a typical feature seen in turbulence simulations, occurs with little influence of beta. This suggests that pressure plays at most a minor role in the early time relaxation processes (see, e.g., \cite{ServidioEA08}).
 
\begin{figure*}
\begin{center}
\includegraphics[width=\textwidth]{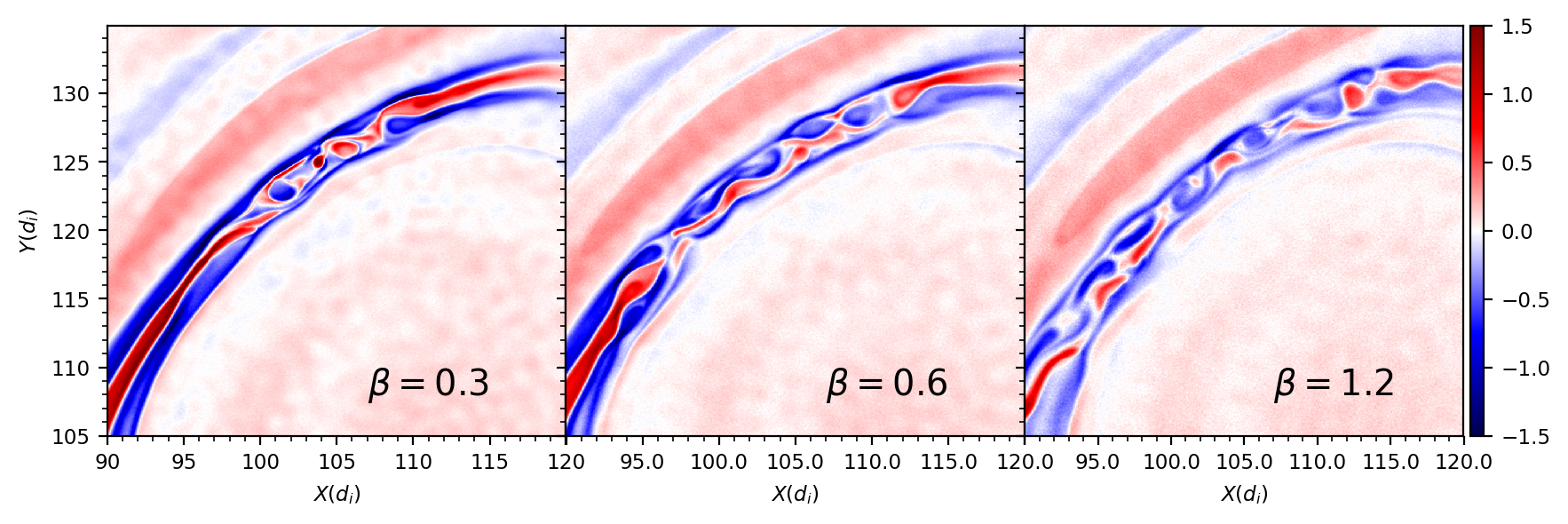}
\caption{(color Online) Out of plane current $J_z$ for the simulations. Only a small part of the domain is plotted to highlight the differences in structure.
 }
\label{cuts}
\end{center}
\end{figure*}

After $E_b$ reaches a maximum value, the decay of the energies begins to show a clear dependence on $\beta$. There is generally less total (flow plus magnetic) fluctuation energy at a given time, for increasing beta. The ordering of the curves shows that the effect of increasing beta is greater for the flow energy than for the magnetic energy. 

Another important aspect of heating is the distribution of internal energy into protons and electrons \cite{CranmerEA09, HowesMNRAS10, GaryApJ12, WuPRL13, MatthaeusEA16, GaryApJ16, HughesApJL17}. Results addressing this issue are summarized in Figure \ref{heating}, again varying $\beta$. Top two panels of fig \ref{heating} show that over time with increasing $\beta$, the protons gain more internal energy, while the electrons gain less internal energy. Therefore the average heating rate of protons increases with $\beta$ while it decreases for electrons. Relative heating is further clarified in the third panel, which shows the ratio $T_i/T_e$ vs time. It is evident that the increasing $\beta$ leads over time to plasma states with higher $T_i/T_e$. Clear trends are also seen when examining the {\it anisotropies}, defined as $A=T_\perp/T_\parallel$ the ratio of perpendicular to parallel temperatures, measured relative to the global mean magnetic field, of the proton and electron temperatures, which are shown in bottom panels as functions of time. At early times, in the initial $\omega_{ci}t<40$ transient relaxation phase, perpendicular anisotropy is favored for both species. However at later times, once the turbulence is more fully developed, anisotropy decreases in time, for every case. However, at any given time, the  ratio $T_\perp/T_\parallel$ for both species is larger for larger $\beta$. Anisotropy for the particular parameters studied in this paper happens to be in the parallel direction. However, as $\beta$ increases fraction of perpendicular heating increases. This indicates stronger perpendicular kicks to both species with increasing $\beta$. 

Increased proton heating indicates more power at proton scales with increasing $\beta$. To quantify this, we begin by examining the out of plane current $J_z$ from all three simulations as shown in figure \ref{cuts}. A small portion of the whole domain is shown to highlight the differences in detail. The overall structure of current in the whole domain is very similar, as can be appreciated even in this small part of the box. However, the current sheets are much more localized for lower $\beta$ and diffuse into broader structures with increasing $\beta$.

\begin{figure}
\begin{center}
\includegraphics[width=3in]{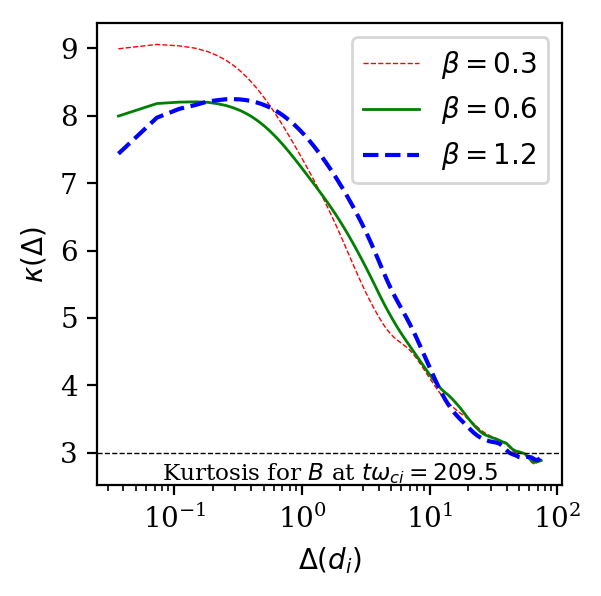}
\caption{Scale dependent kurtosis decreases (shown for magnetic field here) with increasing $\beta$, indicating smaller filling factor and hence weaker strength of intermittent structures.}
\label{sdk}
\end{center}
\end{figure}

\begin{figure}
\begin{center}
\includegraphics[width=3in]{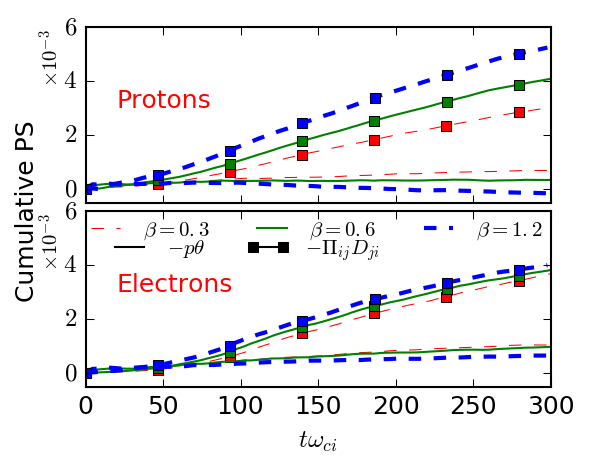}
\caption{Change in dissipation measure, Pi-D as a function of $\beta$. $p\theta$ decreases, and Pi-D increases with $\beta$.}
\label{pid}
\end{center}
\end{figure}

\begin{figure}
\begin{center}
\includegraphics[width=3in]{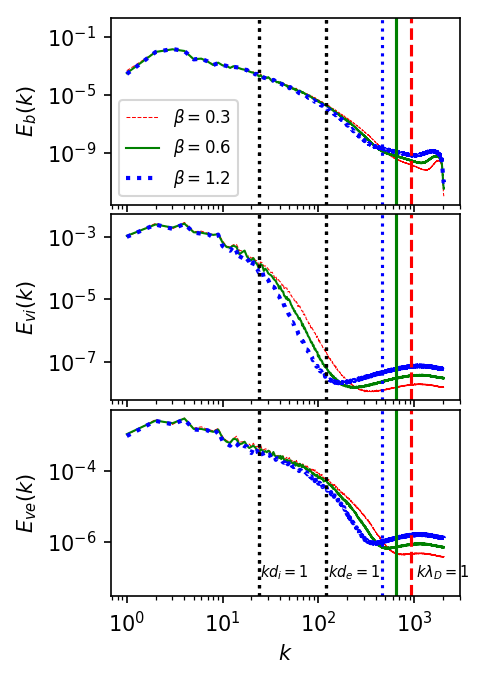}
\caption{Power spectra for magnetic field, proton velocity and electron velocity for various $\beta$. Black dashed lines show proton and electron scales, colored dashed vertical lines show the Debye scale for each simulation, with the colors corresponding to the beta value. $B$ spectra do not show much variation but $\m{v}_i$,$\m{v}_e$ spectra show a clear dependence on beta, with reduced power at appropriate scales for varying $\beta$.}
\label{kspec}
\end{center}
\end{figure}

This broadening of current sheets could imply more intensity at proton scales and lesser intensity at electron scales. This can be easily appreciated in plots of scale dependent kurtosis. In figure \ref{sdk}, we show scale dependent kurtosis for longitudinal increments. We observe that the kurtosis at scales corresponding to $kd_i=1$ (or $\Delta\sim 2\pi$ is the scale of interest) is higher for higher $\beta$ and the kurtosis at electron scales is smaller for higher $\beta$.

The difference in structures with varying $\beta$ also manifests itself
in the pressure-stress dissipation measure ``-PS'' \cite{YangPRE17,
YangEA17ab}, a quantity that measures directly the rate of conversion
of flow energy into internal energy for each species.  Like the ${\bf
J}\cdot {\bf E}$ measure of work done on flows by electromagnetic fields,
the pressure stress interaction is not sign-definite. It is convenient to
separate PS into contributions associated with compressions involving the
dilatation $ \theta = \nabla \cdot {\bf u}$, and a contribution ``Pi-D''
associated with the species traceless velocity stress tensor $D_{ij} =
\partial_iu_j + \partial_ju_i - \delta_{ij}\theta/3$. We therefore write
$-PS =  -p\theta - \Pi_{ij} D_{ij}$ as the total rate of production of
internal energy for a collisionless system.

Figure \ref{pid} shows $p\theta$ and $\Pi - D$ for protons (top panel)
and for electrons (bottom panel). Increasing $\beta$ decreases the
turbulent mach number, and hence the compressibility goes down. This
manifests itself in decreasing contribution from the compressive term
$p-\theta$ for both protons and electrons. However, $Pi-D$ increases
with increasing $\beta$ for both protons and electrons. The change for
protons is, however, more pronounced as compared to electrons.

This clearly shows that ``viscous like'' interactions become more
pronounced with increasing $\beta$ \cite{VasquezApJ12}. Enhanced
viscous-like interactions also reflect their presence in power spectra
for proton and electron bulk flow velocities. Figure \ref{kspec} shows
power spectra for magnetic field ($B$), proton velocity ($\m{v}_i$),
and electron velocity ($\m{v}_e$). The magnetic field spectra do not
show much variability with $\beta$, but the proton and electron velocity
spectra do show variability. Proton velocity spectra show decreased
power at scales between $d_i$ and $d_e$ for larger $\beta$ supporting
the viscous like loss of energy. Electrons show similar decrease in
velocity power at scales between $d_e$ and $\lambda_e$ indicating a
similar mechanism at work for electrons.

\section{Conclusions}
In this study we provide a baseline examination of the effects of
varying plasma $\beta$ on the dynamics of a turbulent collisionless
electron-proton plasma.  To summarize, the results show that
increasing $\beta$ increases the overall decay rate of the fluid scale
fluctuations. Larger $\beta$ also favors proton heating relative
to electron heating.  During fully developed turbulent decay the
temperature anisotropy $T_\perp/T_\parallel$ decreases in time, but it
does so more slowly at higher $\beta$, and in this sense perpendicular
heating is favored at higher $\beta$.  Turning to the pressure-stress
channels of energy conversion, we note that in all cases shown here,
in incompressive, gyro-viscous-like Pi-D channel \cite{YangEA17ab}
is favored over the compressive $p\theta$ channel.  However as $\beta$
is increased the disparity between these two channels increases, for
both protons and electrons.  As Pi-D effects are enhanced, the energy
density in the corresponding velocity spectra is seen to be reduced,
supporting an interpretation of increased viscous-like dissipation
\cite{VasquezApJ12,YangPRE17,DelSartoMNRAS17}.  Meanwhile, increasing
$\beta$ also causes a noticeable change in the coherent structures that
are produced between proton and electron scales.  The structures become
more diffuse at higher $\beta$, a feature that is quantified by the
reduction in the scale dependent kurtosis.

This overall picture is consistent with the elementary idea that
$\beta >1$ implies that pressure effects can overcome magnetic tension
effects. This gives rise to less orderly coherent structures at higher
beta, compared to the magnetically dominated orderly current sheets
and island boundaries that appear in the low beta case. This is the
same idea, applied at kinetic scales, as the rationale provided for
the isotropization of the striae seen in heliospheric imaging as one
passes into the higher beta solar wind from the lower beta corona (see
\cite{DeForestApJ16}). In each case the magnetic field gives up its
previously firm control as beta increases (see also \cite{GhoshPP15-1}
for related discussion).

These present results also show increased proton energization with
increasing $\beta$.  This could be an important factor in explaining
why protons are much hotter than electrons in the magnetosheath. The
magnetosheath is a system that is constantly driven at large amplitudes
by the ram pressure of the solar wind. This produces large amplitude
turbulence in the magnetosheath. This large amplitude turbulence can
heat protons preferentially to electrons \cite{WuPRL13,HughesGRL14,
MatthaeusEA16}. This increases proton beta more than electron beta;
because of this higher $\beta$ protons heat preferentially more than
electrons. The combined effect of large amplitude driving and increasing
$\beta$ can drive protons and electrons to disparate temperature values
until some sort of dynamical steady state is reached.

This preliminary set of results already provides some
guidance in explaining and complementing earlier studies
\cite{WuPRL13,HughesGRL14,MatthaeusEA16}.  (In the final stages
of preparation of this paper, a paper was submitted to arXiv
\cite{KawazuraArXiv18}, exploring some similar questions using a
gyrokinetic-ions and isothermal fluid electron approach. Our results
are consistent with their findings, although the models used are vastly
different). These findings are also consistent with the findings of
\citet{VechApJL17} who find similar trends for proton-electron heating as
a function of turbulent fluctuation amplitude\cite{WuPRL13, MatthaeusEA16,
HughesApJL17} as well plasma beta.  One can readily imagine that more
complex examinations of parameter space will be needed eventually to
answer more detailed questions. For example we have not yet examined the
case of different electron and proton betas, nor of multi-species plasmas.
We have also not considered the effects of collisions or of a population
of neutrals. We leave this for future elaborations on this important
issue on plasma turbulence, namely the role of varying plasma beta.

\begin{acknowledgments}
Authors acknowledge useful discussions with Colby Haggerty at the beginning 
of this project. Research is supported by NSF AGS-1063439, AGS-1156094 (SHINE), 
AGS-1460130 (SHINE),  and NASA grants NNX17AI25G and  NNX14AI63G (Heliophysics Grand Challenge Theory),
the MMS mission through grant NNX14AC39G, and the Solar Probe Plus science team 
(ISIS/SWRI subcontract No. D99031L). We would like to acknowledge high-performance computing support from Cheyenne (doi:10.5065/D6RX99HX) provided by NCAR's Computational and Information Systems Laboratory, sponsored by the National Science Foundation. These simulations were performed as part of the Accelerated Scientific Discovery program (ASD). 
\end{acknowledgments}


\end{document}